\documentclass{article}
\usepackage{amssymb}
\usepackage{amsmath} 
\usepackage{graphicx}

\DeclareMathOperator{\Tr}{Tr}

\title{Mass terms to break susy-like degeneration}

\author{Alejandro Rivero \\ EUPT, Univ de Zaragoza, E-44003 Teruel Spain 
\\ E-mail: {arivero@unizar.es}}

\newcommand{\C}{{\cal{C}}} 

\begin{document}

\maketitle

\abstract{\small We suggest a very simple but general operator to break mass degeneration between representations
of the Poincare group having spin 1 and 1/2. A quantity very similar, at experimental 0.13 $\sigma$ level,
to Weinberg's angle, appears during the process}

\section*{three variations on a same theme}

This preprint is the collection of three separated notes written during the last quarter, aiming to
communicate an amusing finding of a colleague. All the three are different {\it facets} of a same
technical point. We concentrated in this procedure because, as announced in the abstract, it seemed
to have some relationship to the quotient $M_W/M_Z$. It took a long time to us to realize than when the
formulae were adjusted to the mass of $Z^0$ then the value of the electroweak vacuum was also hinted.

\section{Mass terms from Casimir Invariants}
Under Poincare symmetry, suppose we have a family of particles $(m_i,s_i)$ labeled 
using the two Casimirs of the group, $\C_1,\C_2$ with respective eigenvalues
$c_1=m^2$, $c_2=-m^2 s (s+1)$.

We ask for constructions of operators $M^2_s$ with dimension $[mass]^2$ built exclusively from
combinations of this casimirs (excluding inversion) and with the additional asymptotic
condition 
\begin{equation}
\label{cond}
\lim_{s\to \infty} m^2_s = m 
\end{equation}
of recovering the original mass eigenvalue in the high spin limit. This condition allows for 
preservation of the string tension (from the asymptotic Regge trajectory) if for instance
our spectrum of particles comes from a string theory. 

The simplest combination $\alpha \C_1 + \beta \C_2$ of the Casimirs has the adequate dimensions
but fails to meet the asymptotic condition. The next simplest try, and the simplest one
fulfilling our condition, is got from square roots
of the quartic combination. This is, from the solution of the equation

\begin{equation}
M^4_s - M^2_s \C_2 + \C_1 \C_2 =0
\end{equation}

And if we  want to dispose of square roots we must rewrite it in terms
of Pauli Matrices

\begin{equation}
M^2_s=
    \sigma^+ \otimes \C_1 \C_2 + \sigma^- \otimes {\bf I} + 
  { {\bf I} - \sigma_z  \over 2 } \otimes \C_2
\end{equation} 

Note that this operator can be also got from conditions different to (\ref{cond}). An interesting
alternative could be to ask
\begin{equation}
\Tr M_s^2 = \Tr \C_2
\end{equation}

The goal of this note to point out that our method seems to have a role in electroweak
breaking. Meeting with the same equation in a relativistic mechanics context, Hans de Vries
discovered \cite{dVonline} that the positive eigenvalues of this 
operator for s=1/2 and s=1 let one to build the quantity
\begin{equation}
s_{dV}^2\equiv 
1-{m^2_{s=1/2,+}\over m^2_{s=1,+}}=0.22310132...
\end{equation}
unexpectedly near of the mass shell Weinberg angle \cite{pdg,erler}
\begin{equation}
s_W^2=1-{M_W^2 \over M_Z^2} = 0.22306 \pm 0.00033 
\end{equation}

In fact the quotient between de Vries and Weinberg angles 
is $s^2_{dV}/s^2_W=0.9998 \pm .0015$ even too good for a tree level 
prediction, and we should expect it to survive to further experimental updates.

With this ansatz, we can insert the measured $M_Z^2=(91.1874 GeV)^2$ as input for the 
eigenvalue $m_{1,+}^2$ and get the 
other three eigenvalues:

\begin{equation}
m_{s=1/2,+}^2= (80.3717 GeV)^2
\end{equation}

\begin{equation}
m_{s=1,-}^2= - (176.154 GeV) ^2 
\end{equation}

\begin{equation}
m_{s=1/2,-}^2= - (122.384 GeV)^2
\end{equation}

This last negative value is not used in electroweak models, but we find that the negative eigenvalue
$m^2_{1,-}$ is actually in the expected range for the negative mass square operator we use to break
the electroweak symmetry. Remember that 

\begin{equation}
{<v> \over {\sqrt 2}} = \sqrt{-m_h^2 \over \lambda_h} = 174.1042 \ (\pm 0.00075)\ GeV 
\end{equation}

The experimental value coming from Fermi constant \cite{pdg}. So, we are compatible with 
$\lambda_h \approx 1$. In fact we could fix it equal to 1 and pivot on the standard model to
get a tree level estimate of the fine structure constant, getting $\alpha^{-1}=135.28\dots$

It is mysterious why so easily two predictions are got. If we add the actual measurement\cite{top} of the
top yukawa coupling, $\lambda_t=0.991 \pm 0.013$ to our basket and we take it as hint 
for a technicolor/topcolor mechanism, then
one could suspect that techni-forces has also stringy properties --not surprisingly-- and that its
associated string carries somehow a supersymmetry --surprisingly, but a good excuse for $M_W$ to come packed in
a $s=1/2$ object.

\section{A formula to break degeneration of Susy multiplets}

Representations of the 3+1 Poincare algebra can be labeled with two polynomial or Casimir
invariants, ${\cal C}_1$ and ${\cal C}_2$, that in the massive case correspond respectively to the $P^2$
and $W^2$, the latter being the square of Pauli-Lubanski vector. Upon a $(m,s)$ representation
the quadratic Casimir ${\cal C}_1$ has eigenvalue $m^2$ while the quartic Casimir ${\cal C}_2$ has
eigenvalues $-m^2 s (s+1)$.

The goal of this note is to build a new operator of dimension $[mass]^2$ under two restrictions:

1) Use only combinations of the Casimirs, ie the only objects more generally available.

2) Get the same Regge asymptotic trayectory in the limit of high spin. So we request at
least that $\lim_{s\to\infty} M_s=M$, being $M_s$ the new operator.

For a set of equal mass $(m,s_i)$ representations such as the ones happening in a
supersymmetry multiplet, if we want to break mass degeneracy meeting the above conditions 
the simplest way that is to use the formula
\begin{equation}
     M_{(s)}^2 
           \equiv  \frac 12 ( {\cal C}_2 + \sqrt{ ({\cal C}_2)^2 - 4 {\cal C}_1  {\cal C}_2   })
\end{equation}
so that $M^2$ upon a $(m,s)$ representation has eigenvalue
$(m^2/2) (( s (s+1) )^2 + 4 s(s+1))^{1/2} -  s (s+1) )$, that in 
the limit $\it s\to \infty$ approaches to $m^2$. Given
its extreme simplicity this kind of
expressions is not rarely found in  textbooks but we have never
seen suggested its use to break mass degeneracy.

Starting from a primitive relativistic quantum mechanics model,
De Vries found \cite{pf} (see also footnote in \cite{RivVries}) that the 
eigenvalue expression of the above operator, when evaluated both 
at $s=1/2$ and $s=1$ -with degenerated mass- were able to produce a definite number
\begin{equation}
s^2_{dV} \equiv 1 - { M_{s=1/2}^2 \over  M_{s=1}^2}  = 0.22310132...
\end{equation}
and that a mass-related quantity with a similar experimental value seems to exist in Nature;
indeed we can take from the global fit of \cite{pdg,erler}
\begin{equation}
s^2_W=0.22306 \pm 0.00033
\end{equation}

So that the quotient between experimental mass-shell value of Weinberg sine and the theoretical
De Vries "sine" happens to be 
\begin{equation}
s^2_{W,exp}/s^2_{dV} =  0.9998 \pm .0015
\end{equation}

Let us to stress that at the time of De Vries estimate, November 2004, the experimental value and error 
were slightly different so that the $s^2_{dV}$ was more than one sigma away from the measurement. The new results of mass of $W$ and other parameters have moved the global fit so that now $s^2_{dV}$ is very centered inside $0.13 \sigma$.

Of course we have the paradoxical situation that we have produced this quantity in the context of a susy-like relationship between spin 1/2 and spin 1, while Nature seems to have it produced for two spin 1 particles. The transition from one situation to the other shall be given by the still unknown mechanism
of electroweak symmetry breaking. This is to be added to the other mysterious coincidence of the scale of 
electroweak breaking, the value of Yukawian top coupling $y_t$, that currently \cite{top} is expected to
be about $0.991 \pm 0.013$. 
In principle both $y_t$ and $s_W$ are running quantities coming from the GUT scale,
but now we see that they get very singular values just exactly at the moment that the 
electroweak symmetry breaks. 

\section{The $\sin \theta_W$ found in a 1924 timecapsule}

De Broglie's relativistic quantum orbit rule \cite{broglie}
\begin{equation}
\label{eqbroglie}
{m_0 \beta^2 c^2 \over \sqrt {1-\beta^2} }T_r= n h 
\end{equation}
was proposed about the same time that Land\'e-Pauli substitution rule for 3D angular momentum\cite{lande,pauli},
\begin{equation}
\label{eqlande}
{1 \over j^2} \to - {d \over dj } ({1 \over j}) \to {1 \over j} - {1 \over j+1 } \to {1\over j(j+1)}
\end{equation}
but the fast pace of the events in the mid-twenties did not allow for a fusion of both ideas; almost immediately (\ref{eqlande}) was rigorised in the Heisenberg-Born matrix mechanics -- even allowing for half-integer j --, while De Broglie's suggestions for wave mechanics were absorbed into Schr\"odinger's analytic methodology. 

In November of 2004, eighty years later, during an empirical study of gyromagnetic ratios\cite{RivVries}, Hans de Vries suggested to combine (\ref{eqlande}) and (\ref{eqbroglie}) with the extra requirement 
\begin{equation}
\label{eqVries}
 T_r = { h  \over m_0 c^2}
\end{equation}
on the orbital period, so that rest mass and Planck constant are canceled out and we are left with a relationship between relativistic speed and angular momentum:
\begin{equation}
{ \beta^2 \over \sqrt {1-\beta^2} }=  \sqrt {j (j+1)}
\end{equation}

Solving $\beta$ for the $j=1/2$, $j=1$, and via the ratio of speeds, de Vries produced the following adimensional quantity

\begin{equation}
s^2_{dV} \equiv 1 - \big({ \beta_{1/2} \over  \beta_1}\big)^2  = 0.22310132...
\end{equation}
which remembers closely to the mass-based experimental Weinberg's sine.

At the time of calculation the data on $W^+$ mass and the global fits to standard model parameters were putting de Vries' sine at more than $1\sigma$ deviation from the measured value. So the result was put aside as one-line footnote in the preprint report. But the new data released from LEP II during 2005 and the fits from the particle data group have moved the experimental value to be \cite{erler, pdg}
\begin{equation}
s^2_W=0.22306 \pm 0.00033
\end{equation}
so that $s^2_{HdV}$ is now inside the experimental error, centered at $ 0.13 \sigma$. If you prefer, lets say that the quotient $s^2_{W,exp}/s^2_{dV}$ between experimental and theoretical quantities is now $0.9998 \pm .0015$. 

While the experimental error is still too big, the centrality of the calculated result seems to grant that the agreement will continue under further experimental improvements. In any case lets keep in mind that this theoretical number comes from plain relativistic quantum mechanics, thus from the point the view of QFT it is a tree level statement and we should do not expect to push it beyond 0.1\% level; in fact it should be surprising if the experimental error decreases but the central value keeps fixed, because in such case a 0.01\% agreement level would be reached.

De Vries reasonment started from orbital radius instead of orbital period. Indeed one can use the condition (\ref{eqVries}) to get an orbital radius
\begin{equation}
r= \beta \  c \ T_r = \beta  {h \over m c} = \frac h c \frac \beta m
\end{equation}
proportional to Compton length and thus inverse proportional to the orbiting mass. 

Thus we can do the additional remark that if a particle of mass $\propto M_{W^\pm}$orbits according (\ref{eqVries}) producing j=1/2 according (\ref{eqlande})(\ref{eqbroglie}), then a particle of mass $\propto M_{Z^0}$ orbiting at the same radius under the same conditions will produce j=1.

Independently of this remark, we think that model builders can find useful this result. The electroweak scale can be defined as the point at which the renormalised Weinberg's angle, running down from its GUT-theoretical value, reaches the value of de Vries's angle. Besides, de Vries number comes from a pair of well calculated adimensional numbers, 
\begin{eqnarray}
\beta_{1/2}&=\sqrt{\frac 38 (\sqrt{19/3} -1)}=& 0.7541414352817\dots \\
\beta_1&=\sqrt{\sqrt{3}-1}=&  0.855599677167\dots
\end{eqnarray}
 so it contains slightly more information. It could be used for instance to pinpoint mass values at $\beta_{1/2} M_Z \propto \beta_1 M_W \propto 68.76 \ GeV$ or $ M_W/\beta_{1/2} \propto M_Z/\beta_1 \propto 106.5 \ GeV$. Also, the attempt of providing physical meaning to the quotient of speeds (or, via an arbitrary potential, of binding radius) seems to underline composite, top-condensation like, models of the Higgs sector, but we do not put forward a definitive statement on this.


\section*{coda}

Since the redaction of the above notes, I have received a letter from Hans showing that other arrangements can also hit
three digit precision easily, then putting less confidence in a single hitting of a single parameter. I still have some
confidence on the above idea because on one side it has some physical content, as the last note shows, and on another
hand it seems to hit more than one single parameter. Still, for convenience, let me finish reproducing this note from
de Vries, and addressing you towards \cite{dVonline} for detailed comments.

{ \it
I spend some time on other purely numerical coincidences involving the
Weinberg angle, yes, more coincidences...

cos(Theta) =  arcsinh(1)   ---  sW\^{}2 = 0.2231806

This is by far the simplest but it doesn't make so much sense physically,
mW and mZ would be related by some momentum/boost ratio..

The other one is:

sin(Theta)/cos(Theta)  = Beta\_{}1\^{}4  ---   sW\^{}2 = 0.223112151

Where the left term is the ratio in which W3 and B are combined to form
the massless Electromagnetic field in the Weinberg/Salam theory. The
right term is the spin-1 beta 0.85559967716. In correspondence with the
Pauli spinors one could relate  W1,W2,W3 with x,y,z and B with t so the
ratio W3/B could be related to speed, however here we have something to
the power 4....

Well I'm just making a note of them here. Don't know what to do with them.
It made me feel less sure about the one we're using but I still think
that's the one that makes most sense physically.}

\end{document}